# DISTRIBUTED COMPUTING FOR LOCALIZED AND MULTILAYER VISUALIZATIONS

Mark Burgin
Department of Mathematics
University of California, Los Angeles
Los Angeles, CA 90095

Walter Karplus and Damon Liu
Department of Computer Science
University of California, Los Angeles
Los Angels, CA 90095

**ABSTRACT**

The aim of this paper is to develop an approach to visualizations that benefits from distributed computing. Three schemes of process distribution are considered: parallel, pipeline, and expanding pipeline computations. Expanding pipeline structure synthesizes the advantages and traits of both parallel and pipeline computations. In expanding pipeline computing, a novel approach presented in this paper, a multiplicity of processes are concurrently developed in parallel and knotted processor pipelines.

The theoretical foundations for expanding pipeline computing as a computational process are in the domains of alternating Turing machines, molecular computing, and E-machines. Expanding pipeline computing constitutes the development of the conventional pipeline architecture aimed at utilization of implicit parallel structures existing in algorithms. Such structures appear in various kinds of visualization. Image deriving and processing is a field that provides diverse opportunities for utilization of the advantages of distributed computing.

The most relevant to the distributed architecture is stratified visualization with its two cases based on data localization and layer separation. Visualization is treated here as a special case of simulation.

As an example, the concepts developed in this paper have been applied to a visualization in the computer support system utilized in radiology – the noninvasive treatment of brain aneurysms.

# 1. INTRODUCTION

Distributed computing increases efficiency of computer applications but only in the case when the structure of a computational process provides for rational concurrent utilization of different devices. In some cases, problems have an inherent parallelism that implies parallel or, at least, concurrent computations. Other problems are essentially sequential, their solutions cannot utilize the advantages of distributed computing. At the same time, there are many situations when solutions may be realized by different computational schemes. When it is so and multiple devices are accessible, the goal of a programmer or of a software designer is to elaborate such an algorithm that makes more use of the distributed resources.

The aim of this paper is to develop such approaches to visualizations that benefit from distributed computing. Three schemes of such process distribution are considered: parallel, pipeline, and expanding pipeline computations. The third approach, *expanding pipeline computing,* synthesizes the advantages and traits of both parallel and pipeline processing. In expanding pipeline computing, which is a novel approach presented in this paper, a multiplicity of processes are concurrently developed in parallel and knotted processor pipelines.

The theoretical foundations for expanding pipeline computing as a computational process are in the domains of alternating Turing machines (Chandra *et al*. 1981), molecular computing (Sipper 1999), and E-machines (Burgin 1980). Expanding computing constitutes the development of the conventional pipeline architecture aimed at utilization of parallel substructures existing in algorithms. An algorithm as a whole may be not parallel, but some of its parts may express highly parallel behavior. Such structures appear in various kinds of computation. Expanding computing utilizes such peculiarities of computations to achieve higher efficiency by optimizing relation between utilized processors and the speed of computation. Computer visualization is a field that provides diverse opportunities for utilization of the advantages of distributed computing.

An approach that is strongly correlated with distributed computing is *stratified visualization.* In it, image is stratified, i.e., separated into several strata, allowing to compute data for visualization of these strata by different devices (processors). The two natural ways of image stratification are localization and multilayer imaging. *Localization* means that the whole picture is divided into several parts and visualization information for each part is processed on a separate device. *Multilayer approach* presupposes that we have a dynamic picture that is changing in a different mode for some of its components. For example, in animation the whole picture is treated as composed of some actively moving beings (people, animals, etc.) and things (cars, planes, etc.), and the background that changes much slowly. This makes it possible to animate moving things and beings separately from the background. Thus, two strata are utilized. Often, disjoint moving things and different beings are also animated separately, giving birth to many other strata.

Logical means for interpretation and investigation of stratified visualization are provided by the logical theories of possible worlds (Hintikka 1962), which have been formalized by the construction of logical varieties (Burgin 1991; 1995). Under certain conditions, this stratification can considerably enhance the efficiency of computer visualization.

In a more extended concept, visualization is a kind of simulation. Namely, different images in reality as well as other phenomena are simulated (modeled) in computer visualization. Images help researchers and other professionals to understand better what is going on even in those cases when the original system is different from the artificially created images. For example, graphs of functions are visualizations of various types of functions. They are efficiently used in mathematics and its applications. Histograms and scatter diagrams are frequently used by statisticians to achieve better insights into the studied processors and systems. Even more emphasis is made on visualization of practical activities.

Thus, when developing general methods of distributed computing for simulation, we are able to apply them to the problems of visualization. This is why the advantages of distributed computing in the form of expanding pipeline are considered in the context of branching simulation when multiplicity of plausible scenarios are generated. In conventional simulations of complex systems, there arise from time to time uncertainties as to which of two or more alternatives are more likely to be pursued by the system being simulated. Under these conditions the simulationist makes a judicious choice of one of these alternatives and embeds this choice in the simulation model. By contrast, in the branching approach, two or more of such alternatives (or branches) are included in the model and implemented for concurrent computer solution.

As an example, the concepts developed in this paper have been applied to a simulation and visualization task that plays an important role in radiology - the noninvasive treatment of brain aneurysms.

## 2. BRANCHING SIMULATION

To understand better the connections between visualization and simulation, it is useful at first to consider the conventional simulation process and compare it with branching simulation. As described in (Karplus 1992), a scientist who succeeds in getting a simulation to run on a computer and who has gained confidence that it is running properly is like a child with a new toy. Unlike a scale model or a laboratory setup, a simulation is infinitely flexible. A few simple commands entered from a terminal can effect profound changes in the model and the way it transforms inputs into outputs. Here is the opportunity to ask: "What would happen if … ?" and to get an answer in short order … feeling for a while like a master of the universe. In a similar way, engineer uses simulations to consider different options of behavior of the device or software she/he is designing. Each simulation produces some feasible scenarios and each scenario is developed from the very beginning many times.

In contrast to this, branching simulation utilizes more efficiently knowledge obtained in previous runs of the simulation model. Repetitive simulation procedures start not from the very beginning but use common parts of computational routes. For example, let us consider simulation of some process from time $T_B$ to time $T_f$ (Figure 1).

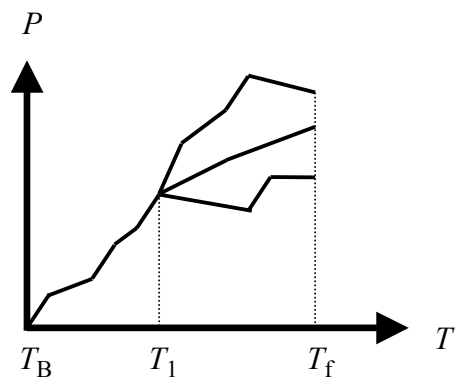

Figure 1: Here $T$ is time and $P$ denotes the process parameters.

In many of such cases, it is desirable to see what happens when at some point $T_1$, which is situated between $T_B$ and $T_f$, the development of the process changes and these changes have many options. It is possible to repeat the whole simulation with different initial data and track all directions of the process development. However, such an approach is far from being optimal. More efficiently, as a rule, is to preserve the results of simulation at the interval [$T_B$, $T_1$] and begin new simulation procedures only from the point $T_1$. Thus, we get branching of processes at this point.

Concurrency and parallelism are urgent demands of modern computations in their search for efficiency. The whole process of branching simulation is intrinsically concurrent (and only in some cases completely parallel). But this concurrency differs from the possible or actual parallelism of conventional simulation processes because each of such processes begins from the very beginning without taking into account all knowledge obtainable in a simulation process. However, in many cases when a new simulation process begins, it is possible to take a part of one of the previous simulations as a part of this new simulation and to realize simulation only for what happens after the chosen part ends. Consequently, it necessitates evaluation of the knowledge obtained in simulation and usage of this evaluation in organizing a directed branching simulation.

Predictions become very unreliable but they may be still useful serving to alert us to the possibility of a catastrophe. However, it is possible to derive even more profit from computer simulation in the case of catastrophes. To do this, it is necessary to consider a new form of prediction when several possible outcomes are predicted in contrast to the conventional prediction when only one "real" phenomenon or situation is predicted. This kind of prediction, we call the *multivariant* prediction. It is consistent with the physical theory of the multi-world universe (Eastbrook 1998). The essence of this theory is that at each moment (at definite moments) of time the universe is divided into several similar but not identical universes which continue to develop in their own directions dividing on every new step of this development. Multivariant prediction can also alert us to what may happen but its main goal is to prepare us to these options.

With respect to the outcome, all predictions are divided into three classes: 1) *descriptive* prediction demonstrating what will be in future; 2) *prescriptive* prediction representing what is necessary to do in future; and 3) *evaluative* prediction explicating what is possible in future.

The multivariant prediction is evaluative by its nature. However, to prepare us, it is necessary for branching prediction to be also prescriptive. Thus we come to a synthesis of evaluative and prescriptive predictions when it is considered what can happen and what is necessary to do in each of possible situations. We call this type the *conditional* prediction. Branching simulation used for prediction is called branching prediction. It makes multivariant prediction more efficient. The salient characteristics in branching approach that distinguish a branching prediction from other predictions is that different branches of simulation that construct feasible scenarios are developed concurrently based on utilization of previous obtained scenarios and systematic evaluation of the events and routes of the development of events.

Theoretical results in the theory of algorithms support these implications. For example, mathematical theory of complexity (Balcazar et al. 1998) makes a distinction between simple and hard problems. Some problems are proved to be very hard because to be solved they demand a lot of time and/or a huge amount of other resources. In theory of complexity, such problems are called NP-complete or PSPACE-complete problems. However, if somebody solves some of these problems in advance and stores the answers in some memory, then finding an answer will take not much time (and other resources). This makes these problems very simple, essentially reducing their complexity.

Logical means for interpretation and investigation of branching simulation and multivariant prediction are provided by the logical theories of possible worlds (Cann 1993), which is formalized by the construction of logical varieties (Burgin 1995). These constructions define semantics for simulation as a whole and for branching simulation, in particular, making possible to apply to them traditional and new logical methodology in researching for simulation processes such problems as realizability, satisfiability, solvability, etc. Many practical problems in programming and computation have formal logical representation, and consequently, they may be treated by logical means that give more founded results than traditional empirical means (Tucker et al. 1994). As an example, we can consider proofs of program correctness.

Simulation may be used to represent future, present or past. In the first case, we call it *prediction.* In the second case, it is called *reflection.* The third case is named *retrospection.* Branching simulation approach is useful not only for prediction, but as well for reflection and retrospection. In this section we consider the case of reflection. The reflection techniques are often exploited by real-time simulations that have been applied to different training and education systems. For example, several simulation systems are developed for the medical specialists that allow individual neurosurgeons or radiologists to learn and visualize complex human anatomy and blood flows, interactively explore established clinical procedures in a computer-generated world, without putting patients at risk. For instance, all of the radiologists using simulating systems can be expected to have similar objectives and exercise only those aspects of the tasks that are relevant to their needs. The kind of interactions that such users will require from the system are therefore predictable to a considerable extent.

There are several ways in which this knowledge of limited number of scenarios can be utilized to maximize the effectiveness of the simulation. To this end, branching might be a promising approach to increase the efficiency of such simulations, making them more adequate to real situations and providing more opportunities to understand the dynamics of the tasks.

However, the prevalent goal in reflective simulation is to present current characteristics and, in many cases, the latest picture of the simulated process. If the process is changed completely in-between two consecutive simulations, then each simulation cycle has to be realized from the very beginning. In many case, only some of all parameters are changing (*the partial parametric stability*), or some parts of the process stay unchanged (*the partial spatial stability*) when the next simulation cycle begins. These features of the process imply that branching can drastically increase efficiency of the simulating system. It is especially essential for interactive real-time systems where time is a critical parameter for simulation. Two types of stability are related to two types of branching: *parametric* and *spatial.*

## 3. PIPELINES IN VISUALIZATION

Real-time computer visualization is one of the most complex and computational intensive problems. Fast, highly parallel computers have been commonly suggested and utilized as a tool to solve such problems.

It is known that the structure of pipelined data flows and computations is well suited for exploring performance issues in visualization. Often, a sequential computer may require minutes to handle the massive amount of computations needed for generating an image frame. By successfully mapping the mathematical recursions of a special class of algorithms into their corresponding computational pipelines, these algorithms can be efficiently carried out by pipeline computers. With this end in view, algorithms which have local or bounded communication structures are preferred.

Typical visualizations involve several types of data represented in multidimensional arrays. Various kinds of operations need to be performed for a time sequence of images, and those operations usually exhibit spatial and/or temporal parallelism, computing with local communications. Pipeline-based computers, such as systolic array processors, therefore offer an opportunity to realize those operations and by that means greatly speed up the computations.

A conventional pipeline computer consists of a set of interconnected processors, each is capable of performing a simple identical task. In each machine cycle, each processor takes values from its input ports, performs the required computations, and passes the results and data to its output ports. Data are used effectively at each stage and the performance analysis describes the importance of keeping the pipeline full.

However, visualizations usually employ algorithms from a wide range of scientific domains, and performance of such algorithms is governed by individual task characteristics. Consequently, issues are important how to design and develop systems that provide for desired flexibility in managing computational resources (load balancing) to implement algorithms for changing task requirements. In the conventional pipeline of processors, it is much more difficult for a visualization system to quickly and flexibly adapt its processing strategy to the actual task characteristics.

## 4. EXPANDING PIPELINE COMPUTING

In the conventional pipeline architecture, computing goes in one direction along a line of processors. Each of these processors perfoms one stage of the complete computational process. To make it possible, processors are connected in a sequence and data are transmitted from one processor to the next (Figure 2a). It is possible to organize several parallel pipelines (Figure 2b).

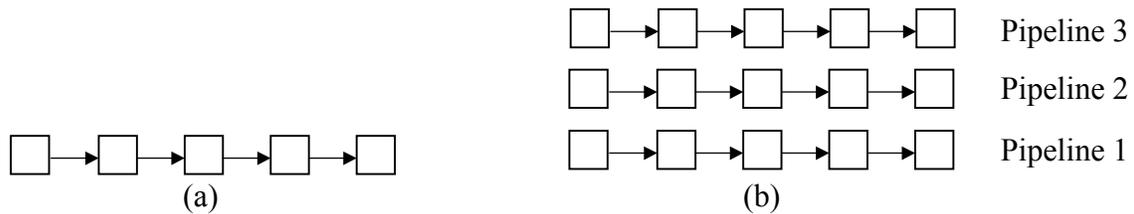

Figure 2: (a) Conventional pipeline architecture; (b) Parallel pipeline array; ▢ denotes a processor; → denotes data transmission.

However, it increases efficiency only when it is possible to decompose the computational process into several independent streams. In contrast to this, computing in the expanding pipeline architecture may generate new computing streams on any level of the pipelines. Consequently, it is necessary to have corresponding pipelines that realize these streams. It demands additional connections between processors (Figure 3a). The most simple case is when a process ramifies at some points after which new process evolves. This new process is realized by another pipeline of processes (Figure 3b). To make utilization of processors in the net more efficient, it is useful to organize adaptive system of processor connections.

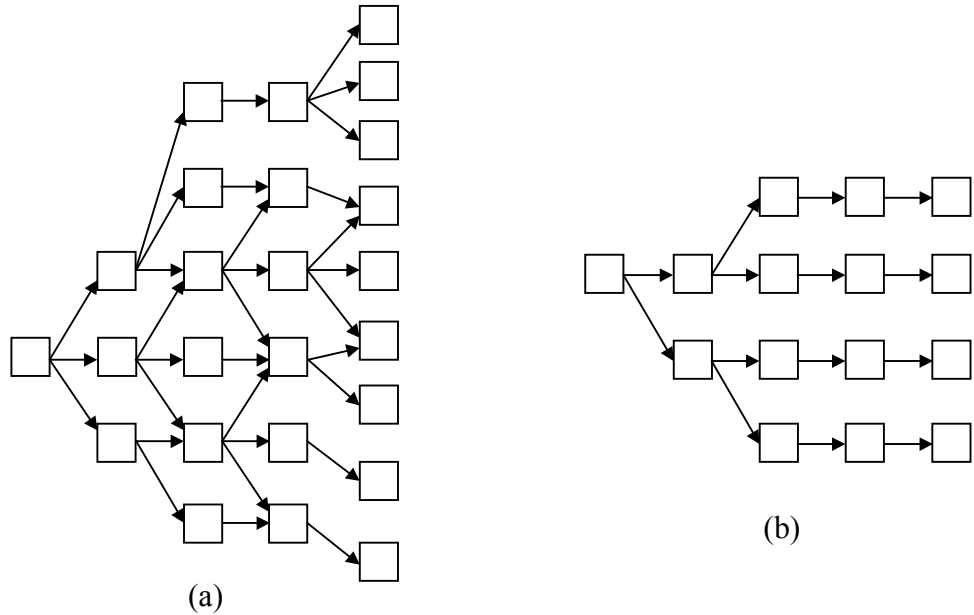

Figure 3: Expanding pipeline array. (a) General case; (b) Semi-parallel case.

To understand better the expanding pipeline approach, let us consider uniform computations with a homogeneous set of initial data *W*. Uniformity here means that the same program *P* is applied to each element *x* from *W*, where *x* is a collection of data for one run of *P*. If it is possible to have access to all data from *W* at the same time, then parallel architecture may be used with a high efficiency. When elements of *W* are computing in a sequence, then it is better to divide *P* into several parts, $P_1, \ldots, P_n$, so that *P* is a sequential composition $P_1, \ldots, P_n$ of these parts. This provides for efficient utilization of the pipeline architecture with $k \leq n$ processors.

However, in many cases such homogeneous sets of data appear on some stage of computation. For example, after we apply program $P_1$ to a collection *x* from *W*, the result of the computation gives birth to several collections of data $v_1, \ldots, v_m$ (m is a natural number) that have to be processed by the program $P_2$. Such situation is especially frequent in the route-finding problems. Starting with the initial point (state), a route-finding program generates a set $V$ of points (states) that may be achieved by one step from the initial point. Then the same procedure is repeated for each point from $V$. Thus, on the first stage the program creates a homogeneous set $V$ of data that afterwards can be processed in a parallel way. Further stages of the route-finding problem give more and more homogeneous sets of data.

To utilize this emerging parallelism, the expanding pipeline architecture has been elaborated. It makes possible to increase productivity of the computing system to a great extend. Theoretical results obtained for alternating algorithms (Chandra et al. 1981) that provide mathematical models for expanding pipeline architecture show this architecture is the most productive (in principle) for all parallel computing architectures. Expanding pipeline architecture is sufficient for realization of the alternating algorithms in the most efficient way. Consequently, it gives an evidence that expanding pipeline approach is more efficient than both conventional pipeline and parallel computing structures. Moreover, the expanding pipeline approach provides the most diverse opportunities for distributed computing.

## 5. VIRTUAL REALITY IN MEDICAL SYSTEMS

In recent years, increased computer usage in clinical medicine and biomedical research has changed the way health care is delivered (Rosen et al. 1996). In particular, the use of virtual reality (VR) technology impacts a number of distinct areas such as: 1) teaching students human anatomy and pathology; 2) surgical procedure training for new surgeons; 3) surgical planning of complex treatments; 4) navigational and information aids during surgery; and 5) predicting operational outcomes.

Technical developments are rapidly improving the computing power and the level of visual realism which provides proper capabilities for the realization of many operative navigation and medical checkups. Such simulations and visualizations seek to augment or enhance the way medicine is taught and delivered by providing a more effective training and education in a safe, controlled, and consistent fashion. In a simulated environment, inappropriate manipulations are acceptable. However, surgeons are expected to learn from such mistakes before progressing to actual patients. Errors should only be made in simulations and never be repeated in real life experiences.

VR also has the potential to revolutionize medical imaging. Three-dimensional medical imaging has been an extremely active field with numerous theoretical, technological and clinical issues. One of the most critical issues is visualization. Three-dimensional visualization is unique as compared with other scientific areas because it has to handle very large volume of data, where the primary limitation is the ability to extract and understand the useful information contents within the real-time constraints, allowing users to interact with realizations of the digitized information itself.

The advances in software, in hardware speed and power, and in computer graphics and visualization technologies are especially helpful in bringing the cost down to more affordable to solve this time bottleneck. Although perhaps only a few years in the future, VR technology will incorporate digital imaging data to give massive improvements in diagnostic treatments.

## 6. BRANCHING SIMULATION IN VA SYSTEM

The methodology of branching simulation in the case of branching reflection has been applied to a simulation and visualization task that plays an important role in radiology - the noninvasive treatment of brain aneurysms. Aneurysm surgery remains dangerous because surgeons have limited knowledge of blood flow patterns and complex anatomy of aneurysms. Part of the problem physicians face is to determine if the aneurysm is suitable for a certain surgical technique. For example, if it is too large, in a different location, or oddly shaped, there may be difficulties. Techniques are often considered in an educational environment with its inherent risks to patients. Accordingly, scientific simulation and visualization of time-varying aneurysm anatomy, pressure, and flow at any point in the vascular system is one example of how VR can help medical specialists roam the complex dataset, evaluate the effects on sealing, blood flow, and pressure of various surgical techniques. Here VR is used for reflection. VR-based systems in medicine represents a paradigm shift in the way that will allow surgeons to evaluate and practice new technologies in a simulator before using them on patients (Chinnock 1994).

At UCLA, researchers are currently building a VR simulation environment to help the neurosurgeon, radiologist, or vascular system specialist plan treatment of brain aneurysm, which is a weakening of the blood vessel walls that can expand like a balloon. Treating aneurysms with implantable coils is designed to seal this expansion to prevent them from bursting (Karplus and Harreld 1994).

The Virtual Aneurysm (VA) system operates on a network that supports client-server paradigm. Flow simulations are computed over time as the heart goes through its pumping cycle. To ensure numerical stability, simulations are computed using small time stepsize such that only a very small fraction of the total data change their values from one step to the next.

Simulations often run for tens or hundreds of hours on high-performance computing machines and periodically generate snapshots of states. The large quantities of simulated data are subsequently stored in archives or databases on disk. After data are off-loaded, they are analyzed and post-processed using VR and scientific visualization techniques to explore the evolving state of the simulated fluid dynamics within the vascular system from local graphics workstations (Liu et al. 1997).

Typical visualizations require a significant amount of I/O bandwidth for accessing data at different time steps when there is not enough memory space for the entire time sequence. The results of data access must be communicated to the graphics workstations, which not only causes significant data movement across slow networks but also interfaces with complex human-computer interactions.

There are two key aspects of computational efficiency of the VA system to which the branching approach can be applied to improve performance: simulation aspect and visualization aspect.

Our Virtual Aneurysm system is based on the numerical solutions of Navier-Stokes equations for the case of three-dimensional time-varying flows. Usually, data describing a mesh geometry is the input, and solution data (velocity, pressure, and so on) at a finite set of points within the computational mesh is the output. The geometry of the mesh is commonly specified by an ordered list of nodes in combination with a list of nodal point coordinates.

Time-varying flow solutions consist of a series of single-time solutions. To ensure numerical stability and desired accuracy of the solutions, it is critical that the time stepsize and the type and distribution of mesh vertices be controlled. Adding this time and mesh control dramatically increases the dataset size, increasing storage requirements and computational complexity.

The physical system modeled is blood flow in a lateral wall aneurysm with its parent artery.

Simulations proceed using small time stepsize and fine grids of mesh elements. Very often, only data values in the regions near or inside the aneurysm have rather noticeable changes, while data values in the other parts of the artery remain static across successive time steps. Many simulated data values at a mesh vertex do not differ markedly from data values at its adjacent vertices in the previous time steps.

Consequently, the branching approach implies many advantages to guide effectively the simulation process. Under program control and without user input or intervention, the revised simulation data at a vertex can therefore re-use the results of the simulations at its adjacent vertices or in the previous time steps, and continue computing only those data values that actually change. In some instances, this unchanging data make up a large percentage of the solution data, allowing simulation time to be saved by avoiding unnecessary or redundant computations. While this method does incur some overheads in time and memory, it still enhances productivity by offering substantial reductions in overall computational and time costs.

## 7. LOCALIZED VISUALIZATION IN VA SYSTEM

Visualization is a simulation of a process with a visual output. The visualization algorithm used to explore CFD data, for example, cutting plane, frequently requires rapid access to subsets of data involves the absolute positions in space. The core technique for loading data into the cutting plane tool is based on a mechanism for returning a value at a given position in space. This position is a 4D space-time point. The value may be a single scalar value like pressure, or a vector quantity like velocity. With a cutting plane, users can view a manageable subset of data and can explore the field by navigating the window through the space.

Flow raw data is generated from a simulation which typically consists of hundreds of time steps worth of information. Due to the spatial and temporal coherence, if two positions are close, the fraction of variation between data is often very small. Our work is aimed at developing a method that exploits the techniques involving the stratified approach, especially in combination with this spatial-temporal properties in the simulated data of consecutive time steps or neighboring locations. The delay in cutting plane visualizations can be eliminated by computing and loading only those data values that have changed in subsequent frames. As a result, the amount of computations and I/O bandwidth required for a subsequent frame can be considerably smaller.

## 8. CONCLUSION

Thus, a new approach to computer visualization is discussed. It is called the *stratified visualization.* In it, image is stratified, i.e., separated into several strata/layers, allowing to compute data for visualization of these strata by different devices (processors). Logical means for interpretation and investigation of stratified visualization are provided by the logical theories of possible worlds, which have been formalized by the construction of logical varieties. Under certain conditions, this stratification can considerably enhance the efficiency of computer visualization.

Stratified visualization provides new opportunities for distributed computing what is urgent when time is a critical resource, a similar advantage has branching simulation for visualization. Branching approach makes possible to spare time of computations in several ways. First, excessive computational branches are determined and eliminated. Excessive computational branches appear due to the fact that some regions of the considered vessels do not change in the process of aneurysm treatment. Second, branching simulation provides conditions for elimination of common subsequences of computational sequences.

Such situations, for example, happen rather frequently in a simulation task that plays an important role in radiology - the noninvasive treatment of brain aneurysms. To cure an aneurysm properly, it is necessary to know fluid characteristics in a blood vessel with the aneurysm. These parameters are provided by the simulation system developed at UCLA. The results of simulation are displayed on a screen to help the doctor who is operating the aneurysm. When the simulation and display programs are used in the interactive mode, time becomes a critical parameter.

A new direction in computing that is beneficial for stratified visualization is cellular computing or its particular case, molecular computing (Adleman 1994; Sipper 1999). As research shows (Sipper 1999), the properties of cellular computing models are flexible and can be tailored to specific tasks. However, it is not an all-encompassing general-purpose approach but a methodology which can excel being applied to the stratified visualization.

Mathematical methodology that provides powerful means for image stratification and distributed processing is the theory of logical varieties (Burgin 1991; 1995). In it each stratta or layer of an image is represented by a separate calculus from the logical variety and interaction between such images goes on inside intersections of these calculi. Event calculi (Shanahan 1999) and logical varieties of event calculi are efficient tools for reasoning about dynamics of visualization processes and for studying their properties.

# REFERENCES

Adleman, I.M. 1994. "Molecular Computation of Solutions to Combinatorial Problems." *Science 266,* no. 5187: 1021-1024.

Balcazar, J.L.; J. Diaz; and J. Gabarro. 1998. *Structural Complexity.* Springer-Verlag.

Burgin, M. 1980. "Some Properties of E-machines." *Abstracts presented to the American Mathematical Society 1,* no. 3.

Burgin, M. 1991. "Logical Methods in Artificial Intelligence Systems," *Notices of the Computer Science Society,* No. 2: 66-78

Burgin, M. 1995. "Logical Tools for Inconsistent Knowledge Systems." *Information: Theory & Applications* 3, no. 10: 13-19

Cann, R. 1993. *Introduction to Formal Semantics.* Cambridge University Press.

Chandra, A.K.; D.C. Kozen; and L.J. Stockmeyer. 1981. "Alternation." *Journal of the ACM 28,* no. 1 (Jan.): 98-108.

Chinnock, C. 1994. "Virtual Reality in Surgery and Medicine." *Hospital Technology Feature Report 13,* no. 18 (Dec.): 38-41, American Hospital Association.

Eastbrook, G. 1998. "What came before Creation?" *U.S. News,* July 20.

Hintikka, J. 1962. *Knowledge and Belief.* Cornell University Press, Ithaca, New York.

Karplus, W.J. 1992. *The Heavens are Falling: The Scientific Prediction of Catastrophes in Our Times.* Plenum Press, New York.

Karplus, W.J. and M.R. Harreld. 1994. "The Role of Virtual Environments in Clinical Medicine: Scientific Visualization." In *Proc. CISS First Joint Conference of International Simulation Societies* (Zurich, Switzerland, Aug. 1994), 13-17.

Liu, D.S.M.; W.J. Karplus; and D.J. Valentino. 1997. "A Framework for the Intelligent Visualization of Large Time-Dependent Flow Datasets in Medical VR Systems." Technical Report No. 970016, UCLA Computer Science Department (May).

Rosen, J.M.; H. Soltanian; R.J. Redett; and D.R. Laub. 1996. "Evolution of Virtual Reality." *IEEE Engineering in Medicine and Biology Magazine 15,* no. 2 (Mar.-Apr.): 16-22.

Shanahan, M. 1999. "The Event Calculus Explained." In A*rtificial Intelligence Today: Recent Trends and Developments,* M.J. Wooldridge and M. Veloso, eds., Springer-Verlag, 409-430.

Sipper, M. 1999. "The Emergence of Cellular Computing." *Computer 32,* no. 7 (July): 18-26.

Tucker, A. et al. 1994. *Fundamentals of Computing I: Logic, Problem Solving, Programs, and Computer.* McGraw-Hill.